\documentclass[preprint, aps, amssymb, pra, onecolumn, longbibliography, unsortedaddress,floatfix]{revtex4-1}
\usepackage{bm}
\usepackage{amsmath}
\usepackage{amssymb}
\usepackage{amsthm}
\usepackage{graphicx}
\usepackage{SIunits}

\usepackage{color}
\usepackage[normalem]{ulem}
\definecolor{blue}{rgb}{0,0,1}

\newcommand{\dd}[1]{}

\begin{document}

\title{Local density of optical states in the band gap of a finite photonic crystal}

\author {Elahe Yeganegi}\email{e.yeganegidastgerdi@utwente.nl}
\author{Ad Lagendijk}
\author{Allard P. Mosk}
\author{Willem L. Vos}
\affiliation{Complex Photonic Systems (COPS), MESA+ Institute for
Nanotechnology, University of Twente, PO Box 217, 7500 AE Enschede, The Netherlands}

\date{\today}

\begin{abstract}
 We study the local density of states (LDOS) in a finite photonic crystal, in particular in the frequency range of the band gap. We propose a new point of view on the band gap, which we consider to be the result of vacuum fluctuations in free space that tunnel in the forbidden range in the crystal. As a result, we arrive at a model for the LDOS that is in two major items modified compared to the well-known expression for infinite crystals. Firstly, we modify the Dirac delta functions to become Lorentzians with a width set by the crystal size. Secondly, building on characterization of the fields versus frequency and position we calculated the fields in the band gap. We start from the fields at the band edges, interpolated in space and position, and incorporating the exponential damping in the band gap. We compare our proposed model to exact calculations in one dimension using the transfer matrix method and find very good agreement. Notably, we find that in finite crystals, the LDOS depends on frequency, on position, and on crystal size, in stark contrast to the well-known results for infinite crystals.

\end{abstract}

\maketitle

 %--------------------------------------Text---------------------------------------------

\section{Introduction}
Many efforts have been stimulated over the years by the theoretical prediction that there exists an infinite three-dimensional, periodic structure that exhibits a three-dimensional (3D) photonic band gap~\cite{Bykov1972spj,John1987PRL,Yablonovitch1987,Soukoulis1996}. In this frequency range the density of optical states (DOS) is zero and vacuum fluctuations are inhibited. This characteristic property of the 3D bandgap crystal leads to novel phenomena in cavity quantum electrodynamics (cQED)~\cite{Bykov1972spj, Yablonovitch1987} where they offer at least five prospects for new physics. Firstly and probably the most eagerly pursued phenomenon is the complete inhibition of spontaneous emission. Any interaction mediated by vacuum fluctuations is affected by their suppression in the bandgap~\cite{Haroche1992}.
Therefore, a crystal with a 3D photonic bandgap not only inhibits spontaneous emission - including a shift of the emitter's frequency known as the Lamb shift~\cite{John1990PRL} - it will also modify the spectrum of blackbody radiation~\cite{Fleming2002N,Fleming2005APL,Babuty2013PRL}, it will affect resonant dipole-dipole interactions including the van der Waals and Casimir forces~\cite{Kurizki1988PRL, Antonoyiannakis1999PRB}, and the well-known F{\"o}rster resonant energy transfer that is prominent in biology and chemistry~\cite{Haroche1992,Novotny2006,Blum2012PRL}. %  
Secondly, once a bandgap is achieved, the cQED physics becomes even richer by introducing point defects.
A point defect acts as a tiny cavity that is shielded in all three dimensions from the vacuum by the surrounding  crystal~\cite{Yablonovitch1991PRL2, Joannopoulos}.
Hence such a photonic bandgap cavity is literally a "nanobox for light".
A third reason why 3D photonic bandgaps are relevant to solid state cQED occurs when a gain medium is introduced in a nanobox.
Such a nanobox with gain offers the promise of a thresholdless laser~\cite{Yablonovitch1987}.
Fourth, an important research theme in cQED is the breaking of the weak-coupling approximation.
There are several ways to break the weak-coupling limit.
One approach is to operate close to a van Hove singularity where the density of states has a cusp~\cite{Ashcroft1976}.
A second approach to break the weak-coupling limit consists of rapidly modulating the "bath" that surrounds a two-level quantum emitter~\cite{Lagendijk1993Lucca}, using ultrafast all-optical switching methods~\cite{Johnson2002PRB}.
Fifth, in quantum physics there is an active interest in decoherence, that is, the loss of coherence between the components of a system that is in a quantum superposition~\cite{Zurek1991PT}.
Hence the shielding of vacuum fluctuations by a 3D photonic bandgap offers opportunities to make optical quantum systems robust to decoherence.

Recently, there have been great attempts to make 3D photonic band gap crystals which show a 3D photonic band gap~\cite{Yablonovitch1991PRL,Ho1994SSC,Wijnhoven1998S,Vlasov2001N,Blanco2000N,Noda2000,
Schilling2005APL,Rinne2007NP,Aoki2008,Staude2012OE,VandenBroek2012}. In practice a real photonic crystal is of course always finite. Therefore the DOS, that is an average over all of space, can not describe the optical states in a \emph{finite} system. For an inhomogeneous system, including finite crystals we need to take into account local properties and therefore we need to calculate the local density of states (LDOS). Having a finite crystal fundamentally changes the main features of the LDOS. First of all the LDOS that is zero in the band gap of an infinite crystal~\cite{Li2001,Busch1998}, becomes non-zero in the band gap of a finite crystal~\cite{Leistikow2011,Fussell2004}. Moreover, within the bandgap of an infinite crystal the LDOS does not depend on frequency or position; in stark contrast in the bandgap of a finite crystal the LDOS does depend on frequency, position, and the crystal size. To date, most theories assume crystals of infinite extent (see Refs.~\cite{Busch1998,Moroz1999,Li2001,Nikolaev2009JOS} and references therein), and can thus not be used to interpret experimental results on real crystals as first studied in Ref.~\cite{Leistikow2011}. To the best of our knowledge the LDOS in the band gap of a finite 3D photonic crystal has not been studied yet as a function of the crystal size and the position in the crystal and there is no analytic theory to calculate the LDOS for any class of crystal. Several groups have calculated optical properties of finite bandgap crystals: Interesting work has been done to calculate the DOS (but not the LDOS) for 1D-waves in the band gap of a finite 1D crystal by Bendickson \emph{et al.}~\cite{Bendickson1996}. The LDOS for 2D waves has been calculated in the band gap of a 2D photonic crystals by Asatryan \emph{et al.} as a function of position, crystal size, and frequency~\cite{Asatryan2001}. The 3D LDOS has been calculated for 1D and 2D crystals that do not have a 3D gap~\cite{Wubs2002,Fussell2004,Prosentsov2007}. 
For 3D light in a 3D inverse opal, Hermann and Hess have calculated the LDOS as a function of frequency, position, and for several position in the crystal by means of finite difference time domain simulations~\cite{Hermann2002}. Unfortunately, however, the results can not be applied to other crystals, physical insights are not readily apparent, as is intrinsic to simulations, and finally such calculations requires a lot of programming efforts and computational costs.
\begin{figure}%[!tbp]
\includegraphics[width=0.5\columnwidth]{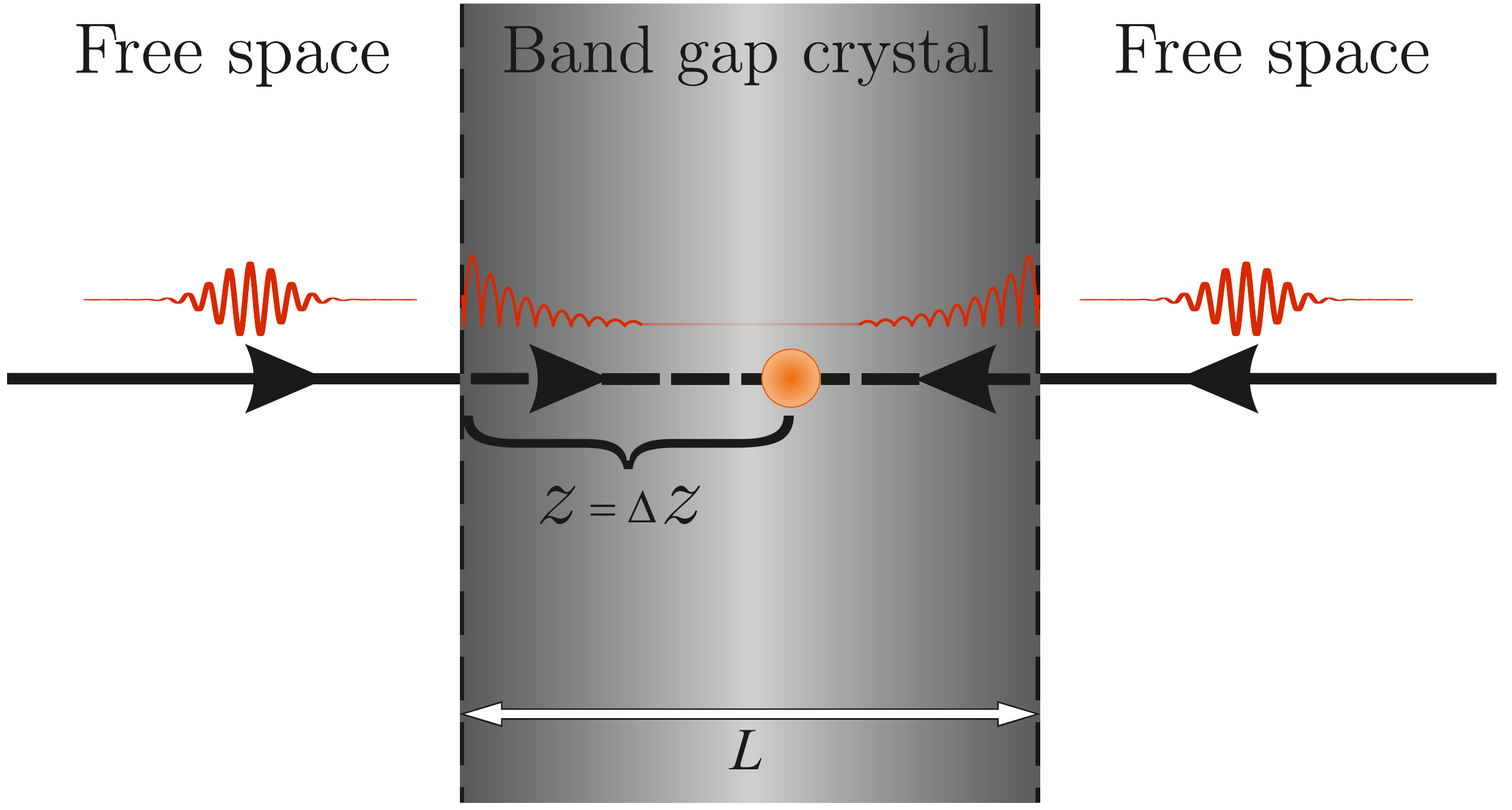}
\caption{
(color) Schematic drawing of a light source (orange sphere) embedded in a finite (1D) photonic band gap crystal with the size $L$ (photonic crystal slab). The source is located at a depth $z= \Delta z$. The interface between the crystal and free space is indicated by the vertical dashed line. The shading in the crystal lightens away from the interface to illustrate the decreasing LDOS with depth into the crystal $\Delta z$. }
\label{emitter_tunneling}
\end{figure}

In this paper we introduce a new point of view on the band gap in a finite photonic crystal, which allows for simplified calculations based on physical principles. We consider a photonic band gap in a finite crystal to be the result of vacuum fluctuations in free space that tunnel into the forbidden zone where the fluctuations are exponentially damped. The physical situation is illustrated in Fig.~\ref{emitter_tunneling} where a light source (orange circle) is embedded in a finite photonic band gap crystal.
While our model is ultimately intended to calculate the LDOS inside the band gap of a finite 2 or 3-dimensional photonic crystal as a function of position in the crystal, crystal size, and frequency, in this paper we will limit ourselves to a 1D structure. The reason is that the electromagnetic field propagation and the band structure in 1D can be calculated analytically, therefore we can test our model by comparing with the analytic calculations.

\section{Local density of states in a finite crystal}

The local density of states (LDOS) $N(\mathbf{r},\omega,\mathbf{e}_d)$ in an infinite photonic crystal is conviniently calculated as a sum over all modes \cite{Sprik1996,Vats2002}:

\begin{eqnarray}\label{eq:LRDOS-general}
N(\omega,\mathbf{r},\mathbf{e}_d) = \frac{1}{(2\pi)^3}\sum_n\int_{BZ}d\mathbf{K}
\delta(\omega-\omega_{n,\mathbf{K}})|\mathbf{e}_{d}\cdot\mathbf{E}_{n,\mathbf{K}}(\mathbf{r})|^2,
\end{eqnarray}
where the LDOS is a function of the emission frequency $\omega$, the source's position in the crystal $\mathbf{r}$, and the dipole orientation $\mathbf{e}_d$.
The integration over the real wave vector \textbf{K} is performed over the first Brillouin zone, and \emph{n} is the band index. $\mathbf{E}_{n,\mathbf{k}}(\mathbf{r})$ is the mode function of the (quantized) electromagnetic field. In a photonic crystal these mode functions are Bloch modes equal to:
\begin{eqnarray}\label{eq:Bloch-mode}
\mathbf{E}_{n,\mathbf{K}}(\mathbf{r})=\bar{\mathbf{E}}_{n,\mathbf{K}}(\mathbf{r})e^{-i\mathbf{K}.\mathbf{r}}.
\end{eqnarray}
where $\bar{\mathbf{E}}_{n,\mathbf{K}}$ has the periodicity of the unit cell. The physical interpretation of Eq.~\ref{eq:LRDOS-general} is that the delta function $\delta(\omega-\omega_{n,\mathbf{K}})$ filters the relevant emission frequency from the collection of photonic states. Since in the band gap of an infinite crystal there are no propagating modes, the delta function causes the LDOS to become zero in this frequency range.

To calculate the LDOS for a finite photonic crystal we modify the LDOS equation for an infinite photonic crystal. We develop a model for the LDOS in a finite crystal which concerns the fields $\mathbf{E}_{n,\mathbf{K}}(\mathbf{r})$ and the modes' linewidths to generalize the $\delta$-function in Eq.~\ref{eq:LRDOS-general}, which are the two main factors in the LDOS.
We propose that fields in the band gap can be calculated using our Ansatz as follows: in the gap the fields are interpolated between the allowed Bloch modes at the band edges that are frequency broadened by finite size and that are multiplied by a frequency dependent exponentially decaying term that describes tunneling. This model describes that at a depth $z$ inside a finite photonic band gap crystal the vacuum modes that enter the crystal in a certain direction $\mathbf{K}$ are exponentially damped. 
The second factor is about the modes in the finite crystal. Modes in a finite crystals are no more delta functions and become broadened due to the finite size of the crystal. We propose to employ Lorentzian functions instead of the delta functions, where the linewidth of the Lorentzian is set by the crystal size $L$. These assumptions are used to calculate LDOS in the band gap of 1D photonic crystal as a function of frequency, depth in the crystal, and the crystal size.

\begin{figure}%[!tbp]
\includegraphics[width=0.5\columnwidth]{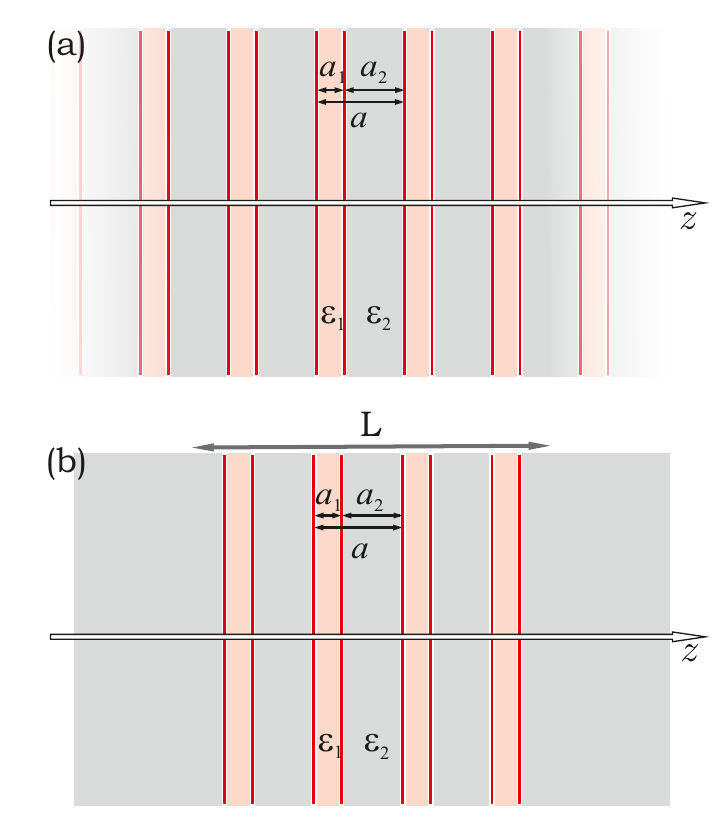}
\caption{
Schematic drawing of a one-dimensional periodic structure. (a) Infinite structure and (b) finite Bragg stack. One period, $a$, consists of two different media with width $``a_{1}"$ and  $``a_{2}"$ and dielectric permittivities of $\epsilon_{1}$ and $\epsilon_{2}$ respectively. The crystal is taken to be one-dimensional and the other directions are used for the illustration purposes.}
\label{Sketch}
\end{figure}

In this paper we describe in detail the above-mentioned viewpoint. In Sec.~\ref{subsec:infinite} we review analytic calculations of the field in a periodic medium. We start with a crystal of infinite extent as is illustrated in Fig.~\ref{Sketch}(a) and we describe Bloch modes as function of frequency and position in the crystal. We also describe the band structure of an infinite 1D photonic crystal in an intuitive way. In Sec.~\ref{subsec:finite} we discuss the field behavior in a finite crystal as a function of frequency, the position and the crystal size. In Sec.~\ref{sec:model} we describe our model for the LDOS in a finite crystal. In Sec.~\ref{sec:checkLDOS} we check our model by calculating the LDOS in the bandgap of a finite photonic crystal using the model and comparing that with the analytically calculated LDOS in the bandgap of a finite 1D crystal. Finally, in Sec.~\ref{sec:conclusion} we summarize our findings.

\section{Field propagation in a periodic media\label{sec:periodicmedia}}
\subsection{Infinitely extended photonic crystal\label{subsec:infinite}}
In this section we investigate the optical properties of an infinitely extended 1D periodic structure as shown in Fig.~\ref{Sketch}(a). To calculate the band structure and the electromagnetic fields we use the transfer matrix method. The details are available in the appendix~\ref{app:ligthpropagation}. 

In Fig.~\ref{B_S} we plot the band structure for a crystal with alternating dielectric constants of $\epsilon_{1}=13$ and $\epsilon_{2}=1$ and layer thicknesses $a_{1}=0.2a$ and $a_{2}=0.8a$. We used $\epsilon_{1}=13$ as an approximation relevant to commonly used semiconductors such as GaAs and Si.
\begin{figure}%[!tbp]
\includegraphics[width=0.6\columnwidth]{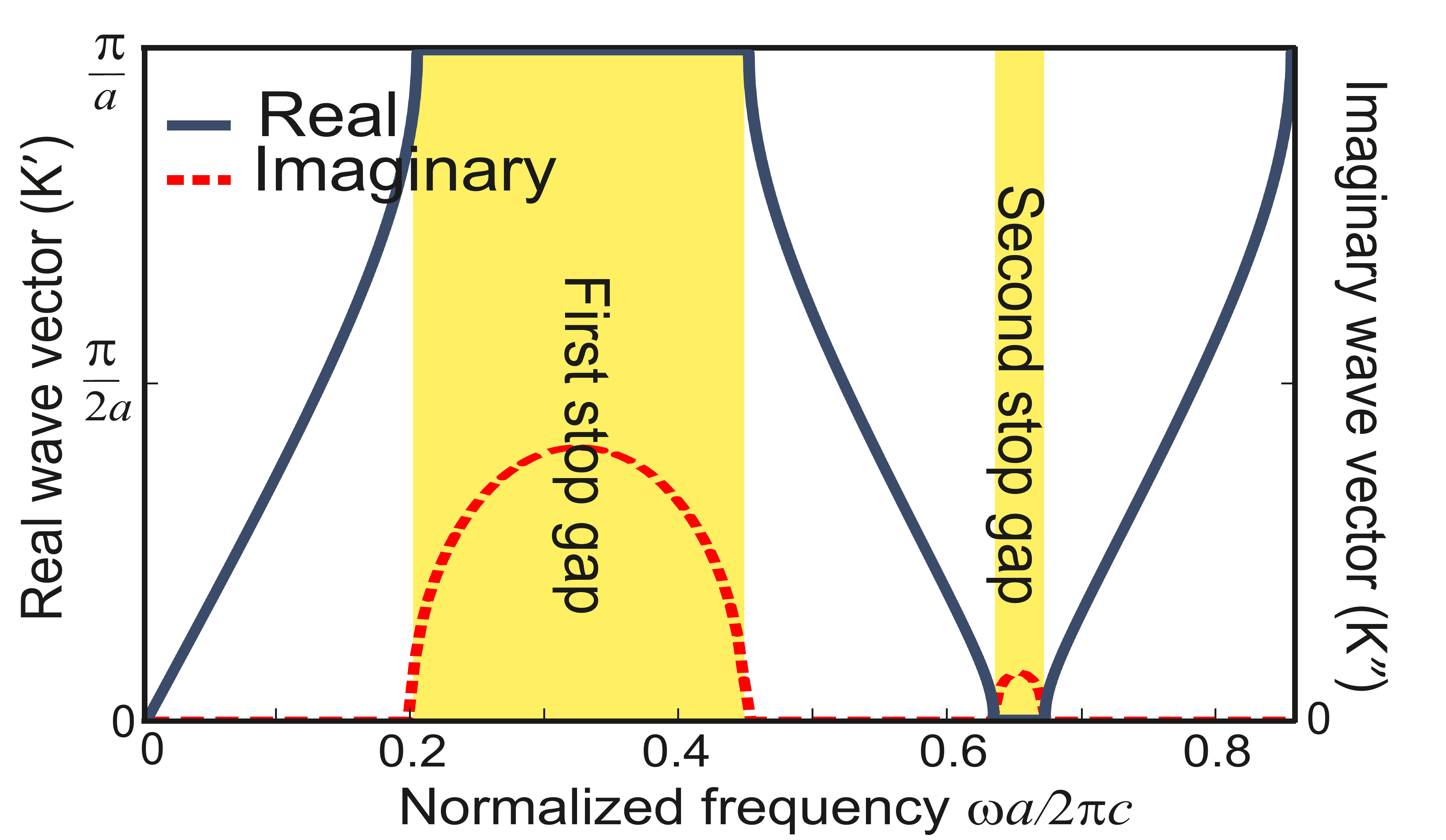}
\caption{
Calculated band structure of the one-dimensional periodic structure depicted at Fig.~\ref{Sketch}, where $a_{1}=0.2a$ and $a_{2}=0.8a$ and $\epsilon_{1}$ and $\epsilon_{2}$ are 13 and 1 respectively. The bands are shown with dark blue lines where the wave vector is real (left ordinate). The first two photonic bandgaps are shown with yellow bars. In the band gaps the wave vector becomes complex $\mathbf{K}=\mathbf{K}^{'}+i\mathbf{K}^{''}$, where the real part is clamped at the Brillouin zone edge wave vector $(K^{'}=\pi/a)$ and the imaginary part $K^{''}$ strongly depends on the frequency as is shown by red dashed line (right ordinate).}
\label{B_S}
\end{figure}
Within the bands $\mathbf{K}$ is real corresponding to the propagating Bloch modes.
In the stopgap, however, $\mathbf{K}$ becomes complex $K=K^{'}+iK^{''}$ and therefore light is attenuated in the crystal.
In this case the real part of the wave vector is clamped at the Brillouin zone edge $K^{'} =\dfrac{\pi}{a}$~\cite{Yariv1983}. Simultaneously the imaginary part strongly depends on frequency $K^{''}=K^{''}(\omega)$, as it increases from zero to a maximum at the stop gap center back to zero at the upper stopgap edge.
\begin{figure}%[!tbp]
\includegraphics[width=1\columnwidth]{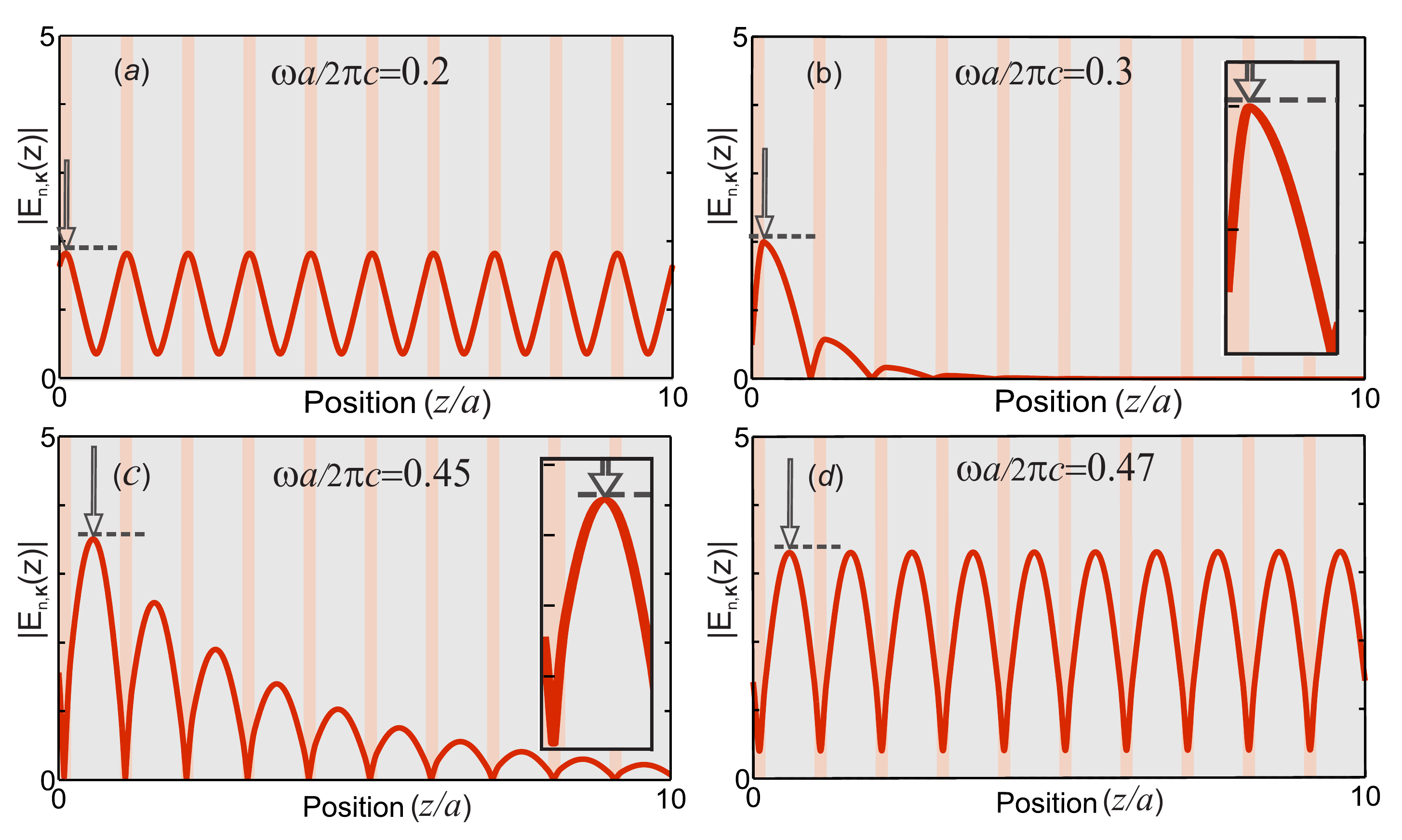}
\caption{
Absolute value of the electric field associated with (a) reduced frequency $\omega a/2\pi c=0.2$, (b) reduced frequency $\omega a/2\pi c=0.3$, (c) reduced frequency $\omega a/2\pi c=0.45$, and (d) reduced frequency $\omega a/2\pi c=0.47$ of the band structure plotted in the Fig.~\ref{B_S}. The main maximum of the field amplitude is indicated in Fig.~\ref{fields} with a black arrow.}
\label{fields}
\end{figure}

Fig.~\ref{fields} (a) shows the calculated absolute value of the field $\vert \mathbf{E}_{n,\mathbf{K}}(z)\vert$ in the first band at a reduced frequency $\omega a/2\pi c=0.2$. It is seen that the field maxima are located in the material with the high dielectric permittivity. This general and well-known behavior occurs at all frequencies in the first band up to the band edge at $\omega a/2\pi c= 0.203$ \cite{Joannopoulos,Yariv1983} . If we assume light to be inside such an infinite structure at frequencies in the band gap, the field is exponentially damped since the incident light is reflected from each layer due to Bragg diffraction.  
Fig.~\ref{fields}(b) and (c) show fields at two different frequencies in the band gap, where Fig.~\ref{fields}(b) is for a frequency near the center of the band gap and Fig.~\ref{fields}(c) corresponds to a frequency close to the band edge where the imaginary part $K''$ of the wave vector is much smaller than at the center of the gap. Fig.~\ref{fields} (d) shows the field at a frequency in the second band where field propagates in the structure and the field maxima are located in the material with the low $\epsilon$, in agreement with the common lore for ``\emph{air bands}'' \cite{Joannopoulos,Russell1995}.
\begin{figure}%[!tbp]
\includegraphics[width=0.55\columnwidth]{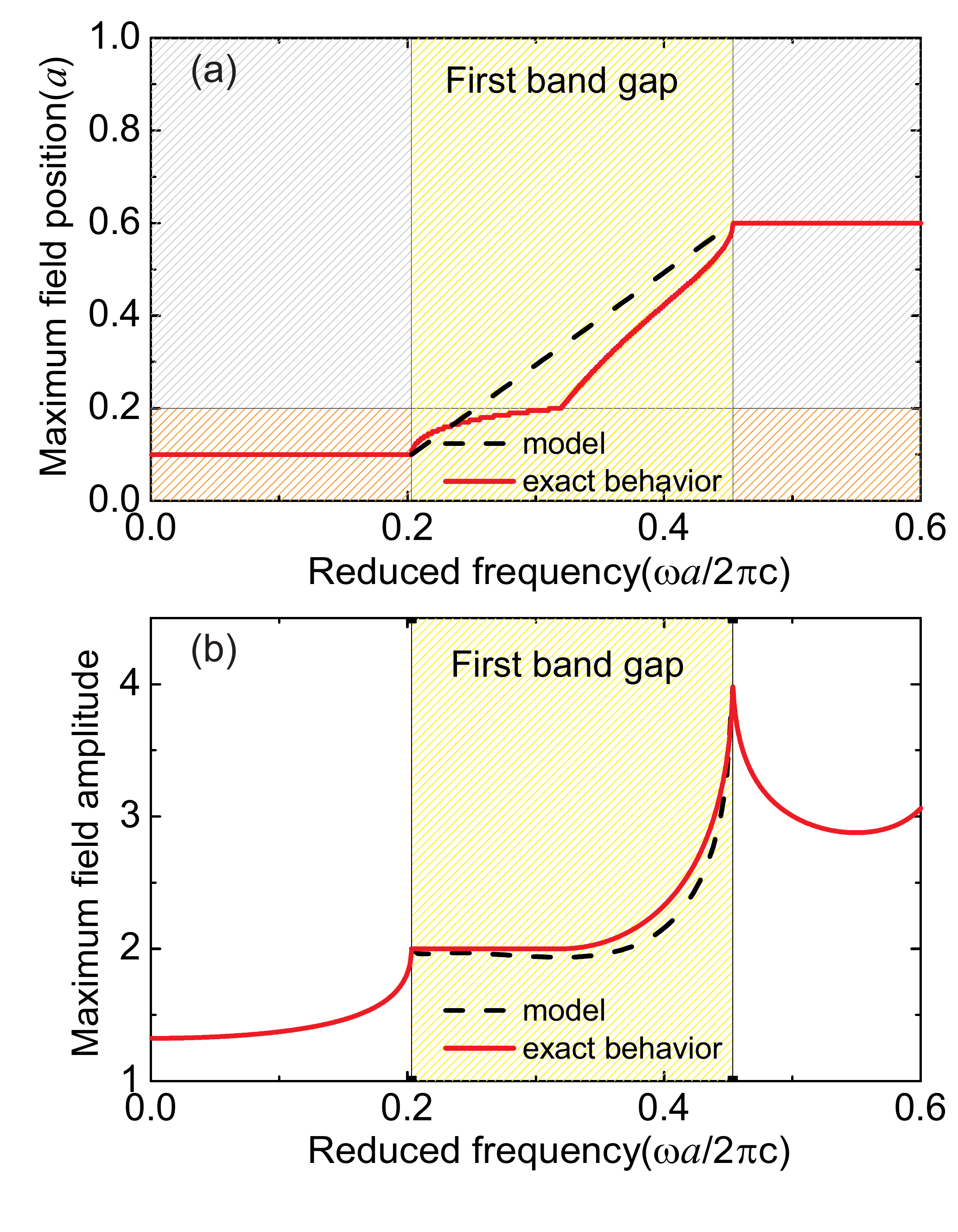}
\caption{
(a)Behavior of the first field maximum in the frequency range of the first bandgap. The ordinate between $0$ and $0.2$ is the high index material (indicated by the orange hatched bar) and from $0.2$ to $1$ is the low index material (indicated by the gray hatched bar). The red solid line is the exact behavior of the field and the black dashed line is the linear field interpolation with the position.
(b) Comparison of the exact maximum field amplitude (red curve) and the model (black triangles) as a function of reduced frequency. In the model a linear interpolation is used for the position of the field maximum (see(a)). The model shows a very good agreement with the exact calculations.
}
\label{E_behavior}
\end{figure}

Now we turn to the field behavior as a function of frequency and position in the periodic structure. To build the model for the LDOS in a finite crystal we consider the main maximum of the field amplitude for different frequencies inside and outside the band gap.
Fig.~\ref{E_behavior}(a) shows the position of the field maximum within a unit cell as a function of normalized frequency. The ordinate between $0$ and $0.2$ is the high index material and from $0.2$ to $1$ is the low index material. The abscissa shows the frequency range, which includes the first and second band and the first band gap which is indicated with the yellow bar. In the first band the field maximum is in the middle of the high index material. In the second band the field maximum is in the middle of the low index material. This field behavior for propagating bands is well-known and corresponds to the dielectric band and air band nomenclature that pertains to a simple photonic crystal, see Ref~\cite{Joannopoulos}. In the band gap, however, the field maximum moves continuously from the middle of the high index layer to the middle of the low index material. It is seen in Fig.~\ref{E_behavior}(a) that this trend through the unit cell is a complex function of frequency. From separate calculations we find that the path is a complicated function of $a_{1}$, $a_{2}$, $\epsilon_{1}$ and $\epsilon_{2}$. In the band gap, the field amplitude $\vert \mathbf{E}_{n,\mathbf{K}}(z)\vert$ does not fulfill the symmetry of the crystal anymore -- here in 1D mirror symmetry at the center of either layer 1 or layer 2 -- since the fields are no more \emph{bonafide} eigenfunctions of the crystal. The trajectory of the field maximum through the unit cell is a phenomena known in X-ray literature, where it is used in X-ray fluorescence standing wave spectroscopy~\cite{McMorrow2011}.

Fig.~\ref{E_behavior}(b) shows the maximum field amplitude as a function of frequency. In the first band the field amplitude maximum increases with frequency until it reaches the lower edge of the band gap. While the trend of the field maximum is continuous in moving from the first band to the band gap and then to the second propagating bands, there is a cusp at each band edge. At the frequency of the upper band edge the field amplitude is enhanced. The maximum of the field amplitude in the second band is higher compared to the first band. This enhancement can be understood from the conservation of field energy $\vert \epsilon \mathbf{E}_{n,\mathbf{K}}(z)^2 \vert$, since the field maximum in the second band is centered in the low index material. This detailed investigation of the field behavior in the band gap suggests a relation between the fields in the gap and the fields at the band edges.

\subsection{Finite photonic crystal\label{subsec:finite}}
The advantage of a 1D calculation is that analytic calculations of the field and of the LDOS for a finite crystal is possible. In this section we calculate the field and the LDOS for a finite periodic structure that has the same periodicity and dielectric materials as the infinite one. We first calculate the field at different frequencies outside and inside the band gap and then using these fields we calculate LDOS inside the band gap of a finite structure. Later on in Sec.\ref{sec:checkLDOS} we compare the LDOS calculated using our model with the exact LDOS calculated using transfer matrix method.

To calculate the field in a finite structure we also use a transfer matrix method. The main difference between the calculations in a finite and infinite structure is that in the infinite structure fields are Bloch modes and therefore we have to apply an extra condition to imply field periodicity as a result of Bloch conditions. This reduces the number of initial conditions for the wave that propagates into the crystal. In case of the finite structure, the Bloch condition does not pertain once we have the freedom to choose any arbitrary initial conditions. Here we choose the incident field in vacuum to have an amplitude equal to unity, irrespective of frequency:
\begin{equation}
\mathbf{E}(z)= b e^{i\mathbf{k}z},  
\label{eq:field_fin}
\end{equation}
\\  with $b=1$.
\begin{figure}%[!tbp]
\includegraphics[width=0.6\columnwidth]{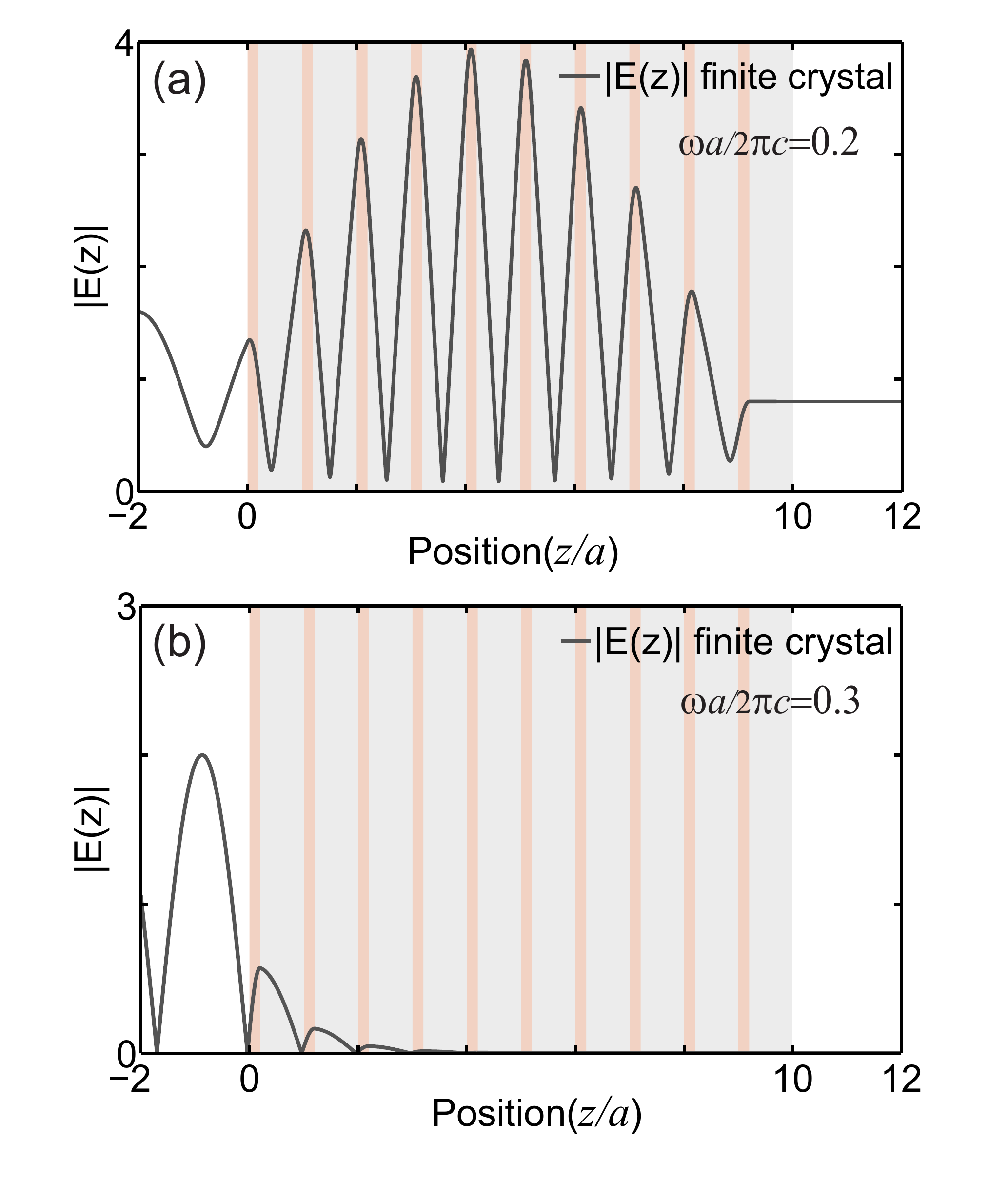}
\caption{
(a) Absolute value of the electric field at a frequency $\omega a/2\pi c=0.2$ for a finite structure with $m=10$ periods. The environment of the crystal is assumed to be air. (b) Absolute value of the electric field at a frequency $\omega a/2\pi c=0.3$ inside the photonic band gap frequency (see Fig.~\ref{B_S}). At this frequency the field inside the structure is damped. The field in the finite structure decays with the same damping as the field in the infinite structure at the same frequency.}
\label{compare_fields}
\end{figure}
Fig.~\ref{compare_fields}(a) shows the exact analytic calculations of the absolute field for a finite crystal consists of $m=10$ layers. The environment is assumed to be free space. At a reduced frequency $\omega a/2\pi c=0.2$ outside the band gap, the field is propagating through the crystal. As seen in the figure the field is not periodic and clearly differs from the field in the infinite structure at the same frequency (see Fig.~\ref{fields}) as a result of Fabry-Perot resonances due to the front and back surfaces.
Fig.~\ref{compare_fields}(b) shows the electric field for the same structure at a reduced frequency $\omega a/2\pi c=0.3$. This frequency is in the bandgap and here the field is damped. In the finite structure the field is damped with the same decay length as the field in the infinite structure at the same frequency, as is expected.

\begin{figure}%[!tbp]
\includegraphics[width=0.6\columnwidth]{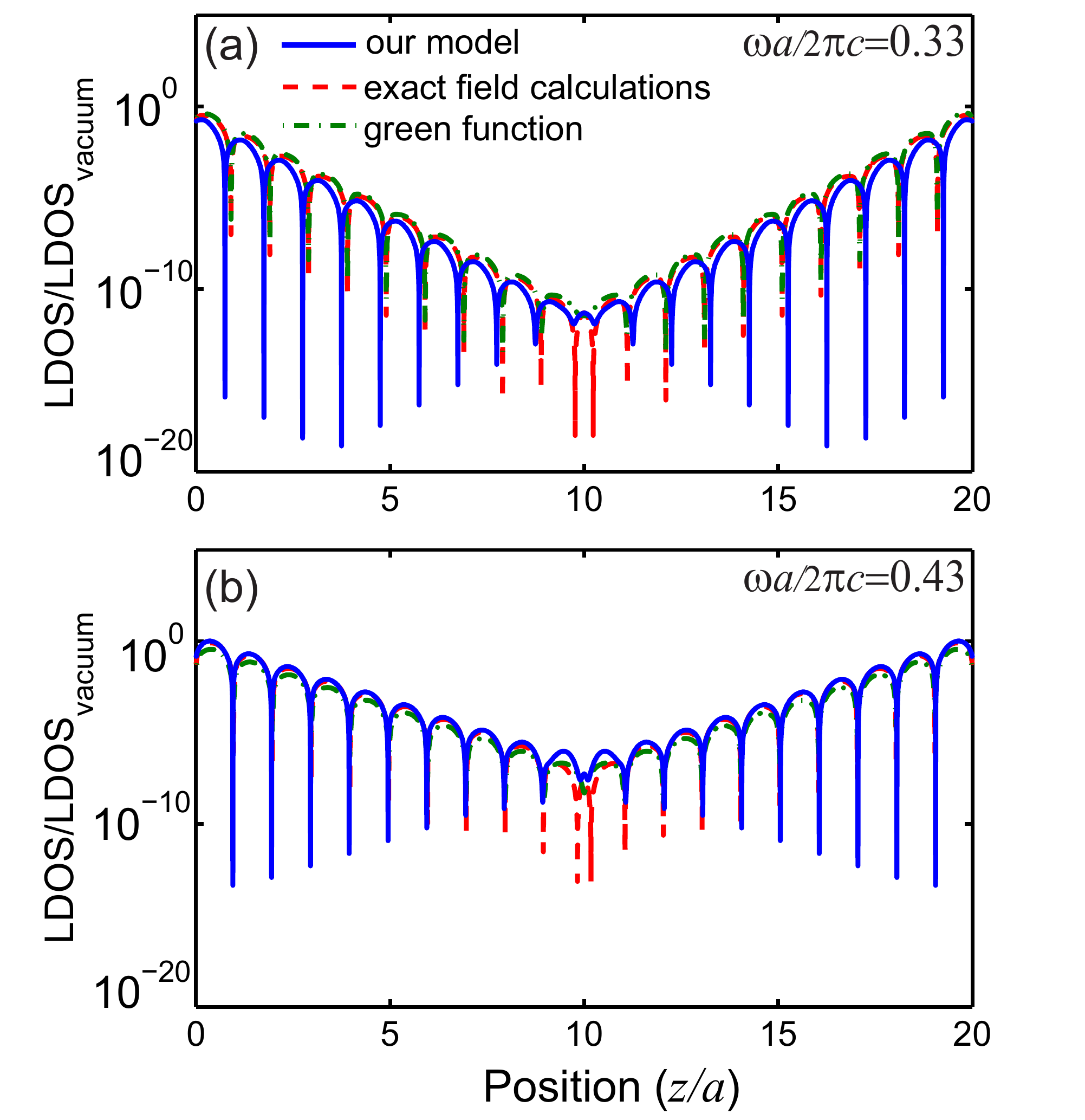}
\caption{ (color)
The LDOS calculated using the Ansatz~Eq.~\ref{eq:Ansatz_} (blue) and the exact LDOS (red) calculated for a finite periodic structure at two different reduced frequencies $\omega a/2\pi c=0.33$ and $\omega a/2\pi c=0.43$ within the photonic bandgap (logarithmic axis). At frequency $\omega a/2\pi c=0.33$ there is a small shift in the unitcell modulations of the LDOS. At frequency $\omega a/2\pi c=0.43$ our model matches perfectly with the exact calculation. This small deviation between our model and the exact calculations is due to our simple interpolation and can be already seen in \ref{E_behavior}(b) at the same frequency.}
\label{LDOS_finite_model}
\end{figure}

To calculate the LDOS we have calculated the field entering the crystal from both sides. Since the dipole orientation of the light source is in the direction of the field propagation, therefore the intensity of the light entering the crystal is directly proportional to the LDOS. In Fig.~\ref{LDOS_finite_model} we show the calculated LDOS in the band gap of the finite crystal. The crystal has a larger size of $m=20$ unit cells and has the same dielectric constants and layer thicknesses as before. We also calculate LDOS using the Green function where we assume that the emitter is at position ($z$) in a host layer~\cite{Smolka2010,Thyrrestrup2012} . The Green function can be solved self-consistently at any position in the structure. The imaginary part of the Green function leads to the LDOS that is shown in Fig.~\ref{LDOS_finite_model} by green dot-dashed line~\cite{Smolka2010} which agrees well with the LDOS calculation using the field. It is seen that in the band gap the LDOS decreases exponentially with depth $z$ in the crystal. At a frequency near the center of the bandgap as in Fig.~\ref{LDOS_finite_model}(a) the LDOS decays faster compared to a frequency more close to the band edge as in Fig.~\ref{LDOS_finite_model}(b), as is expected due to the smaller magnitude of the imaginary wave vector $\mathbf{K}''$ near the edge (see~Fig.\ref{B_S}). The LDOS behavior in the band gap will be discussed in more detail in Sec.~\ref{sec:checkLDOS}.

\section{Model for the LDOS in a finite crystal\label{sec:model}}
To calculate the LDOS in the band gap of a finite crystal we implement our new point of view which consists of two main items. Firstly in a finite crystal, the modes broaden and they have a finite width, therefore in the band gap we see the effect of them. $F(\omega_i,\mathbf{K}_j)$ describes how modes broaden in frequency due to the finite size of the crystal
\begin{eqnarray}\label{eq:Lorentzian}
F(\omega_i,\mathbf{K}_j) = \frac{1}{(2\pi)} \frac{\Delta}{(\frac{\Delta}{2})^2+(\omega-\omega_{i,j})^2},
\end{eqnarray}
where $\Delta$ is width of the Lorenzians that is taken to be the same for every Lorentzian function as it is taken to be determined only by the ratio of the lattice parameter to the crystal size $L$
\begin{eqnarray}\label{eq:Lorentzian_width}
\Delta = \frac{a}{L}.
\end{eqnarray}
Due to the finite size of the crystal the modes become discrete~\cite{Ashcroft1976} and we have a summation of the Lorentzian modes. $\omega_{i,j}$ is the central frequency of each mode and therefore the central frequency of each Lorentzian function. $m$ is the number of wave vectors in the Brillouin zone which is equal to the number of the unit cells in the finite crystal, which stems from the Born-Von Karman boundary condition in solid state physics~\cite{Ashcroft1976}. $L$ is the crystal size that is equal to $L=m\cdot a$. The number of Lorentzian functions is thus set by number of modes in the finite crystal and is therefore equal to $m$ for each band. The expression for the Lorentzian functions $F(\omega_i,\mathbf{K}_j)$ leads to delta functions for an infinitely large crystal $(L\longrightarrow\infty)$, in agreement with Eq.~\ref{eq:LRDOS-general}.

The second item describes the waves in the band gap. To calculate the waves in the band gap we use two features: the Bloch modes at the edges of the band gap, and the imaginary part of the wave vector $\mathbf{K''}$. As we have seen in Sec.\ref{sec:periodicmedia}, in the bands at frequencies below and above the gap the wave vectors are completely real: $\mathbf{K}=\mathbf{K'}$ and $\mathbf{K''}=0$ and the waves are propagating. In the bandgap the wave vectors become complex $\mathbf{K}=\mathbf{K'}+i\mathbf{K''}$ thus the waves are exponentially damped as a result of Bragg diffraction. Here we apply our new point of view to the waves in the band gap. We propose to write the waves in the bandgap as a Bloch function with a complex wave vector as follows:
\begin{align}
\mathbf{E}^{pbg}_{n,\mathbf{K}}(\mathbf{r}) & =\bar{\mathbf{E}}_{n,\mathbf{K}}(\mathbf{r})e^{i\mathbf{K}\cdot \mathbf{r}} \nonumber \\ 
& =\bar{\mathbf{E}}_{n,\mathbf{K}}(\mathbf{r})e^{i(\mathbf{K'}+i\mathbf{K''})\cdot \mathbf{r}}\nonumber \\
& =\bar{\mathbf{E}}_{n,\mathbf{K}}(\mathbf{r})e^{i\mathbf{K'}\cdot \mathbf{r}} e^{-\mathbf{K''}\cdot \mathbf{r}}\nonumber \\
& =\hat{\mathbf{E}}_{n,\mathbf{K}}(\mathbf{r})e^{-\mathbf{K''}\cdot \mathbf{r}}
\label{eq:Ansatz_}
\end{align}
where the mode function $\hat{\mathbf{E}}_{n,\mathbf{K}}(\mathbf{r})$ is periodic at frequencies outside the gap. In the gap the fields are interpolated between the allowed Bloch modes at the band edges. This model describes that at a depth $\mathbf{r}$ inside a finite photonic band gap crystal the vacuum modes that enter the crystal in a certain direction $\mathbf{K}$ are exponentially damped by a factor $e^{-\mathbf{K''}\cdot \mathbf{r}}$, where $\mathbf{K''}$ is the imaginary wave vector for propagating in the direction with unit vector $\mathbf{K}$ and is considered to be positive. 

The black dashed line in the Fig.~\ref{E_behavior} shows how we interpolate the field inside the band gap of a 1D photonic crystal. We know that the fields at the band edges are propagating. Here the absolute value of the field has the same period as the crystal. In the band gap we use a linear interpolation of the position of the field maximum as a function of $z$ and therefore of $\omega$ to interpolate the periodic part of the mode function into the gap. The field amplitude is also linearly interpolated from its value at the lower edge to the upper edge. While a linear interpolation of the position is not an exact description of the behavior shown in Fig.~\ref{E_behavior}, we have chosen it as it is a simple type of interpolation, and therefore a robust one. Nevertheless, the resulting maximum field amplitude matches well with the exact result. When in future more insight is obtained in the behavior of fields in an \textit{n}-dimensional band gap crystal, the interpolation can hopefully be improved. By this interpolation we have obtained the periodic part of the field at each frequency $\hat{\mathbf{E}}_{n,\mathbf{K}}(z)$. To obtain the complete field $\mathbf{E}^{pbg}_{n,\mathbf{K}}(z)$ we then multiply the interpolated field $\hat{\mathbf{E}}_{n,\mathbf{K}}(z)$ with the corresponding decaying factor $e^{-\mathbf{K''}z}$ at each frequency where $\mathbf{K''}(\omega)$ is obtained from the band structure calculations (see Fig.~\ref{B_S}). 
The physical situation that we considered is shown in Fig.~\ref{emitter_tunneling}. A light source is embedded in a finite photonic crystal. 
Using our Ansatz that was introduced at Eq.~\ref{eq:Ansatz_} and inserting it in Eq.~\ref{eq:LRDOS-general} we obtain a general expression for the LDOS in a finite n-dimensional photonic band gap crystal:
\begin{eqnarray}\label{eq:LRDOS-PC-damping}
N^{pbg}(\omega,\mathbf{r},\mathbf{e}_d) = \frac{1}{(2\pi)^3}\frac{1}{m}\sum_{i=1}^{n}\sum_{j=1}^{m}
F(\omega_i,\mathbf{K}_j)\cdot\\
 |\mathbf{e}_{d}\cdot \hat{\mathbf{E}}_{n,\mathbf{K'}}(\mathbf{r})|^2 \nonumber 
\cdot exp(-2 \cdot \mathbf{r} \cdot \mathbf{K''})
\end{eqnarray}
where $F(\omega)$ is the function that describes how the modes broaden in a finite photonic crystal.

In Eq.~\ref{eq:LRDOS-PC-damping} we have distinguished the integration over real wave vectors $\mathbf{K'}$ from damping by imaginary wave vectors $\mathbf{K''}$, and the factor 2 in the exponential originates from the mode functions squared in the LDOS. Thus at frequencies in the photonic band gap a light source at position $\mathbf{r}$ in a photonic band gap crystal experiences a continuum of vacuum fluctuations, all exponentially damped depending on their directions. 
%At frequencies outside the band gap, Eq.~\ref{eq:LRDOS-PC-damping} reverts to the usual expression Eq.~\ref{eq:LRDOS-general} since the imaginary wave vectors then vanish, hence the additional exponents all equal 1. 
The expression Eq.~\ref{eq:LRDOS-PC-damping} leads to the correct limit of perfect inhibition ($N^{pbg} = 0$) for an infinitely large band gap crystal - consistent with Eq.~\ref{eq:LRDOS-general} - since a light source is then infinitely far away ($\mid \mathbf{r}\mid\longrightarrow\infty$) from the surface.

\section{LDOS in the band gap of a finite 1-D periodic structure\label{sec:checkLDOS}}

In this case the integral in Eq.~\ref{eq:LRDOS-PC-damping} reverts to a simpler summation. 
We note that in this case the dipole orientation $\mathbf{e}_{d}$ is in the same direction as $\hat{\mathbf{E}}_{n,\mathbf{K'}}$.

\begin{figure}%[!tbp]
\includegraphics[width=0.85\columnwidth]{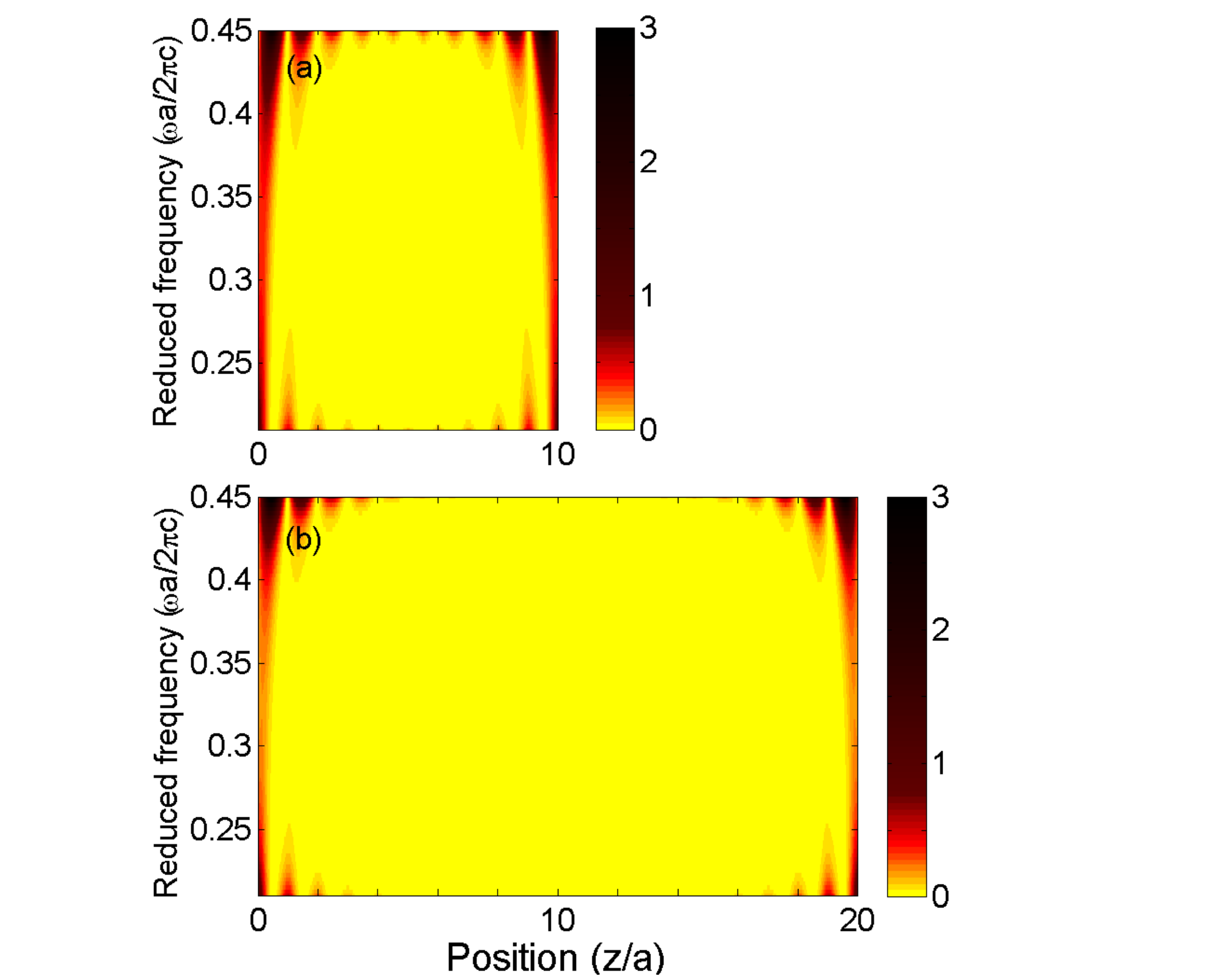}
\caption{
Local density of optical states (LDOS) in the frequency range of the band gap for (a) a finite crystal consisting of $m=10$ periods and (b) a finite crystal consisting of $m=20$ periods. In the band gap the envelope of the LDOS decays exponentially inside the structure with a minimum at the center of the structure. The attenuation of the LDOS is strongest at the center of the band gap because the imaginary part of the $K''$ vector is maximal. Moreover the summation over all $F(\omega_i,\mathbf{K}_j)$ is minimal at frequencies in the center of the band gap.}
\label{LDOS_2ta}
\end{figure}

Fig.~\ref{LDOS_2ta} shows the calculated LDOS in the band gap using Eq.~\ref{eq:LRDOS-PC-damping} for two different crystal sizes of $m=10$ and $m=20$ unit cells. It is apparent from the figure that the LDOS is non-zero in the band gap of both finite crystals, where it is zero in the band gap of an infinite crystal. Furthermore the LDOS becomes position dependent, frequency dependent and crystal size dependent, in contrast to the infinite crystal case where the LDOS does not depend on position, frequency, and crystal size. It is seen that LDOS is modulated in the unit cell due to interferences between the reflections from different layers in the crystal. This spatial dependence of the LDOS in the gap of a finite crystal is in stark contrast with the independence of position in an infinite crystal. In addition, the LDOS decays exponentially as a function of position into the crystal and has a minimum near the center of the structure. The exponential decay is faster at frequencies at the center of the band gap as the damping rate depend on the magnitude of the imaginary part of the wave vector $\mathbf{K''}$. Beside this fast decay at the center of the gap the LDOS magnitude gets also smaller since the amplitude of the Lorenzian modes gets smaller in the middle of the band gap. 

Fig.~\ref{LDOS_2ta} (a) and (b) allow us to compare the LDOS for two different crystal sizes. The LDOS in the central layer of the larger crystal with $m=20$ layers is lower compared to the crystal with $m=10$ layers. This illustrates that if we have a light source in the center of a large crystal with a large number of unit cells, the inhibition of spontaneous emission is higher due to a better shielding of vacuum fluctuations than in a small crystal with a few unit cells.

To verify our model for the LDOS in the band gap of a finite photonic band gap crystal, we compare the LDOS calculated using our model with the exact calculation of the LDOS in a finite 1D periodic structure. Fig.~\ref{LDOS_finite_model} shows both exact LDOS calculations using transfer matrix method for a 1D, $m=20$ periods structure and the LDOS calculated using our model for the same crystal at the same frequency. It is apparent in Fig.~\ref{LDOS_finite_model} that the result from the calculated LDOS using our Ansatz is in very good agreement with the exact LDOS. In both cases LDOS shows the unit-cell modulations. The exponential decay in the LDOS at the same frequency is also the same and all the curves are in good mutual agreement.

\section{Conclusion\label{sec:conclusion}}
We have developed a new model to calculate the LDOS in the band gap of a finite photonic crystal. This model considers the broadening of the electromagnetic modes due to the finite size of the crystal. Our model also introduces an interpolation to calculate the electromagnetic field in the band gap using the band structure and the electromagnetic field at the band edge. A comparison with exact, analytical calculations for 1-dimensional periodic structures verifies the model.
% We propose that this model can be used to calculate LDOS as a function of frequency, crystal size, and at various positions in a finite 2 and 3-dimensional photonic crystal. 
From our results we have gained a better overview of the field behavior and especially the LDOS behavior in the band gap of a finite periodic structure. We find that in the band gap of a finite structure the LDOS depends on frequency, position, and on crystal size, in contrast to infinite structure where the LDOS is zero everywhere.

\appendix
\section{Light propagation in periodic media\label{app:ligthpropagation}}
In this section we describe the optical properties of an infinitely extended 1D periodic structure. To calculate the band structure and the electromagnetic fields we have used the transfer matrix method~\cite{Yariv1983,Pendry1991}. 
The dielectric function of the structure is real and periodic: 
\begin{equation}
\epsilon(z)=\epsilon(z+a),
\end{equation}
where $a$ is an arbitrary lattice vector~\cite{Ashcroft1976}. This means that the structure is invariant under the translation $( m\cdot a )$ where $m$ is an integer and $a$ is the period of the structure. Here we limit ourselves to propagation of light in the $z$-direction perpendicular to the layers, and we assume that the medium is non-magnetic. In this 1D periodic structure it is only necessary to consider one polarization due to the symmetry. The propagation of light with frequency $\omega$ in a periodic structure is described by Maxwell's equations~\cite{Joannopoulos,Yariv1983}. Due to symmetry these equations remain the same after substituting $(z+a)$ for $z$ in the operators and $\epsilon$. The solutions of the wave equation have the form of a plane wave times a function with the periodicity of the lattice
\begin{equation}
\mathbf{E}_{n,\mathbf{K}}(z)=\bar{\mathbf{E}}_{n,\mathbf{K}}(z)e^{-i\mathbf{K}\cdot z},
%H(z)=H_{K}(z)e^{-iK\cdot z},
\label{eq:bloch1}
\end{equation} 
where $\bar{\mathbf{E}}_{n,\mathbf{K}}(z)$ is periodic with the lattice 
\begin{equation}
\bar{\mathbf{E}}_{n,\mathbf{K}}(z)=\bar{\mathbf{E}}_{n,\mathbf{K}}(z+a),
%H_{K}(z)=H_{K}(z+a),
\label{eq:bloch2}
\end{equation} 
which is known as the Bloch-Floquet theorem. The subscript $K$ indicates that the function $\bar{\mathbf{E}}_{n,\mathbf{K}}(z)$ depends on the Bloch wave number $K$. Note that Eq.~\ref{eq:bloch1} and Eq.~\ref{eq:bloch2} imply that:
\begin{equation}
\mathbf{E}_{n,\mathbf{K}}(z)=\mathbf{E}_{n,\mathbf{K}}(z+a)e^{-i\mathbf{K}\cdot a},
%H(z)=H(z+a)e^{-iK.a},
\label{eq:bloch_condition}
\end{equation} 
The problem at hand is thus of determining $K$ and $\bar{\mathbf{E}}_{n,\mathbf{K}}(z)$ as a function of $\omega$. To this end we have used the transfer matrix method together with the Bloch condition to calculate the band structure and the Bloch fields propagating in the structure. The electric field within each homogeneous layer is expressed as a sum of an incident and a reflected plane wave:

\begin{equation}
\mathbf{E}_{n,\mathbf{K}}(z)=b^{-}e^{-i\mathbf{k}z}+b^{+}e^{i\mathbf{k}z},
\label{eq:efield}
\end{equation}
where $k=\frac{n \omega}{c}$ is the wave vector and $b^{-}$ and $b^{+}$ are coefficients that are related through the continuity conditions at the interfaces. 
Using Maxwell's equations one obtains the general form of the magnetic field amplitude:
\begin{equation}
\mathbf{B}_{n,\mathbf{K}}(z)=b^{-}\sqrt{\epsilon} e^{-i\mathbf{k}z}-b^{+}\sqrt{\epsilon} e^{i\mathbf{k}z},
\end{equation}
Imposing the continuity of the electric and magnetic fields at two consecutive interfaces, leads to a matrix that connects the coefficients of one homogeneous layer($n$) to the coefficients of the same type of layer($n+1$) in the next period. These can be written as the following matrix equation:
\begin{equation}
 \begin{pmatrix}
  b_{n-1}^{-}  \\
  b_{n-1}^{+} 
 \end{pmatrix}=
 \begin{pmatrix}
  A & B  \\
  C & D 
 \end{pmatrix}\begin{pmatrix}
  b_{n}^{-}  \\
  b_{n}^{+} 
 \end{pmatrix}
 \label{eq:transfermtx}
\end{equation}
 where $ T= \begin{pmatrix}
  A & B  \\
  C & D 
 \end{pmatrix} $ is a transfer matrix that relates the coefficients of the same type of layers. The matrix elements $A$, $B$, $C$, and $D$ are complex functions of the layers properties such as the dielectric permittivity ($\epsilon_{i}$) and the width ($a_{i}$) of each layer. 
 %Note that there the transfer matrix which relates the field amplitude in the second layer is different from the $T$ matrix above.
 As a consequence, only the first two components $b^{-}_{0}$ and $b^{+}_{0}$ (or the first column vector $ \begin{pmatrix}
  b_{0}^{-}  \\
  b_{0}^{+} 
 \end{pmatrix} $) can be chosen arbitrarily. If we choose the column vector of layer 1 in the zeroth unit cell ($n=0$), then the remaining column vectors of the equivalent layers are related to that of zeroth unit cell by :
  \begin{equation}
 \begin{pmatrix}
  b_{n}^{-}  \\
  b_{n}^{+} 
 \end{pmatrix}=
 \begin{pmatrix}
  D & -B  \\
  -C & A 
 \end{pmatrix}^n\begin{pmatrix}
  b_{0}^{-}  \\
  b_{0}^{+} 
 \end{pmatrix}
\end{equation}
Besides the continuity conditions that should be satisfied, the periodic layers must be invariant under lattice translation, in other words the Bloch condition Eq.~\ref{eq:bloch_condition} must be satisfied.
In terms of our column vector representation and from Eq.~\ref{eq:efield}, the Bloch condition Eq.~\ref{eq:bloch_condition} yields
  \begin{equation}
 \begin{pmatrix}
  b_{n}^{-}  \\
  b_{n}^{+} 
 \end{pmatrix}=
 e^{-iKa}
 \begin{pmatrix}
  b_{n-1}^{-}  \\
  b_{n-1}^{+} 
 \end{pmatrix}
 \label{eq:blochbondry}
\end{equation}
By combining Eq.~\ref{eq:transfermtx} and Eq.~\ref{eq:blochbondry} the column vector of the Bloch wave satisfies the following eigenvalue equation:
\begin{equation}
 \begin{pmatrix}
  A & B  \\
  C & D 
 \end{pmatrix}\begin{pmatrix}
  b_{n}^{-}  \\
  b_{n}^{+} 
 \end{pmatrix}=
 e^{iKa}
 \begin{pmatrix}
  b_{n}^{-}  \\
  b_{n}^{+} 
 \end{pmatrix}.
 \label{eq:eigenequation}
\end{equation}
The eigenvalues are equal to 
\begin{subequations}
\begin{equation}
e^{iKa}=\frac{1}{2}(A+D)+ [\frac{1}{4}(A+D)^2-1]^{1/2}
\label{eq:eigenvalue1}
\end{equation}
\begin{equation}
e^{iKa}=\frac{1}{2}(A+D)- [\frac{1}{4}(A+D)^2-1]^{1/2}
\label{eq:eigenvalue2}
\end{equation}
\end{subequations}
that are the reciprocal of each other. The eigenvectors corresponding to the eigenvalues are
\begin{equation}
\begin{pmatrix}
  b_{0}^{-}  \\
  b_{0}^{+} 
 \end{pmatrix}=
 \begin{pmatrix}
  B  \\
  e^{iKa}-A 
 \end{pmatrix}
 \label{eq:initialcondition}
\end{equation}
This bring us to the point that although for the transfer matrix method one column vector $ \begin{pmatrix}
  b_{0}^{-}  \\
  b_{0}^{+} 
 \end{pmatrix} $ can be independently chosen, since we have imposed the Bloch condition this column vector will be set by Bloch condition as Eq.~\ref{eq:initialcondition}.
That means that depending on $K$ there exists a unique initial condition for which the electromagnetic field propagates into the periodic structure and this initial condition depends on the properties and the geometry of the periodic structure.
To normalize the initial condition we have normalized the field such that if we consider both layers to be vacuum the intensity of the field propagating in vacuum is equal to one and does not depend on frequency, therefore:
 \begin{equation}
\begin{pmatrix}
  b_{0}^{-}  \\
  b_{0}^{+} 
 \end{pmatrix}=
 \begin{pmatrix}
  1  \\
  (e^{iKa}-A)/B 
 \end{pmatrix}
 \label{eq:initialconditionnorm}
\end{equation}
According to Eq.~\ref{eq:blochbondry} the corresponding field amplitudes for the $n^{th}$ unit cell are equal to
\begin{equation}
 \begin{pmatrix}
  b_{n}^{-}  \\
  b_{n}^{+} 
 \end{pmatrix}=
 e^{-inKa}
 \begin{pmatrix}
  1 \\
 (e^{iKa}-A)/B
 \end{pmatrix}
\end{equation} 
With the initial conditions and the Bloch wave vector $K$ we obtain the fields and dispersion relation between $\omega$ and $K$ for the Bloch wave function.
\begin{equation}
K(\omega)=\frac{1}{a}cos^{-1}[\frac{1}{2}(A+D)].
\end{equation} 

\section*{VI. Acknowledgments}
We thank Henri Thyrrestrup and Femius Koenderink for discussions. We thank Stephan Smolka, Henri Thyrrestrup, and Peter Lodahl for making their 1-D Green function code available for us. This work is part of research program ``Plasmonics'' of the Stichting voor Fundamenteel Onderzoek der Materie (FOM) that is financially supported by the Nederlandse Organisatie voor Wetenschappelijk Onderzoek (NWO). We also thank STW and ERC grant 279248 for support.

%\bibliographystyle{unsrt}
%\bibliographystyle{prsty}
%\bibliographystyle{apsrev}
%\bibliography{manuscript_1D_calculation}

\end{document}